\documentclass[twocolumn,showpacs,aps,amsmath,amssymb,superscriptaddress,prl]{revtex4-1}
\usepackage{amsfonts}
\usepackage{amsmath}
\usepackage{bm}
\usepackage{graphicx}
\usepackage{overpic}
\usepackage{color}
\usepackage{multirow}
\usepackage[normalem]{ulem}

\newcommand{\be}{\begin{equation}}
\newcommand{\ee}{\end{equation}}
\newcommand{\bea}{\begin{eqnarray}}
\newcommand{\eea}{\end{eqnarray}}

\newcommand{\Fig}[1]{Fig.~\ref{#1}}

\newcommand{\Eq}[1]{Eq.\,(\ref{#1})}

\newcommand{\Eqs}[1]{Eqs.\,(\ref{#1})}

\newcommand{\w}{\omega}
\newcommand{\ep}{\epsilon}

\begin{document}

\title{Kondo cooling in quantum impurity systems}

\author{Xiangzhong~Zeng}

\affiliation{Department of Chemical Physics, 
University of Science and Technology of China, Hefei, Anhui 230026, China}

\author{Lyuzhou~Ye} 

\affiliation{Department of Chemical Physics, 
University of Science and Technology of China, Hefei, Anhui 230026, China}


\author{Long~Cao}

\affiliation{Department of Chemical Physics, 
University of Science and Technology of China, Hefei, Anhui 230026, China}

\author{Rui-Xue~Xu}

\affiliation{Department of Chemical Physics, 
University of Science and Technology of China, Hefei, Anhui 230026, China}


\author{Xiao~Zheng} \email{xz58@ustc.edu.cn}

\affiliation{Department of Chemical Physics, 
University of Science and Technology of China, Hefei, Anhui 230026, China}


\author{Massimiliano Di Ventra} \email{diventra@physics.ucsd.edu}
\affiliation{Department of Physics, University of California, San Diego, La Jolla, California 92093, USA}

\date{Submitted on March~15, 2022}

\begin{abstract}

The Peltier effect is the reverse phenomenon of the Seebeck effect,
and has been observed experimentally in nanoscale junctions. 
However, despite its promising applications in local cooling of nanoelectronic devices, the role 
of strong electron correlations on such a phenomenon is still unclear.
Here, by analyzing the thermoelectric properties of 
quantum impurity systems out of equilibrium, 
we unveil the essential role of electron-electron interactions
and quantum resonant states in Peltier cooling, leading to the prediction of the {\it Kondo cooling} phenomenon.
The existence of such Kondo cooling is validated 
by a reverse heat current and a lowered local temperature in a model junction. 
The discovery of this unconventional Peltier cooling offers a new approach
toward nano-refrigeration, and highlights the unique role  
of strong electron correlations in nonequilibrium quantum systems.

\end{abstract}

\maketitle

{\it Introduction.}-- 
The Peltier effect is a fundamental thermoelectric phenomenon.  
It means that when an electric current is passed through a junction connecting two conductors (A and B), 
heat will be transferred from A to B, which may lead to cooling in A. 
%
%
Peltier refrigerators \cite{giazotto2006opportunities,Muhonen2012Micrometre} have been designed and applied to cool various types of electronic devices \cite{Rowe1983Modern,nolas2001thermoelectrics,phelan2002current,rowe2006thermoelectrics,bakker2010interplay,Flipse2012Direct}.
This is very appealing for practical purposes, 
because otherwise 
the local heat generation and accumulation will likely 
deteriorate the performances of, or cause damages to the electronic devices.

In the past decade, Peltier cooling in nanoscopic systems 
has attracted enormous theoretical \cite{d2006local,huang2007local,galperin2009cooling,liu2011effect,dubi2011colloquium,finch2009giant,bergfield2010giant,karlstrom2011increasing,santhanam2016thermal,saffarzadeh2018thermoelectric}
and experimental \cite{grosse2011nanoscale,vera2016direct,cui2018peltier,jin2018exploring,uchida2018observation,das2019systematic,wang2020magneto,zhang2020exploring,hu2020enhanced,hubbard2020electron} efforts.  
Thanks to the impressive advances in the synthesis, fabrication and manipulation of
nano-sized materials, Peltier cooling has been realized in 
molecular junctions \cite{grosse2011nanoscale,vera2016direct,cui2018peltier,hu2020enhanced,hubbard2020electron},
organic thermoelectric flakes \cite{jin2018exploring,zhang2020exploring},
and ferromagnetic/ferrimagnetic metal films \cite{uchida2018observation,das2019systematic,wang2020magneto}. 
For instance, Cui {\it et~al.} have 
directly observed the 
Peltier cooling in gold junctions with 
conjugated molecules, and also revealed the relationship between cooling and 
electron transport characteristics \cite{cui2018peltier}. 

Despite the remarkable progress, 
the strength of cooling 
in nanojunctions has not met the needs of practical applications.
For instance, the measured cooling power in Ref.~[\onlinecite{cui2018peltier}]
is lower than $1$\,nanowatt, thus leaving plenty of room for improvement. 
%
Theoretically, the Peltier effect in nanojunctions has been explored
within the Landauer theory of quantum transport \cite{landauer1957spatial,buttiker1985generalized,frensley1990boundary,meir1992landauer,Meir1993Low,jauho1994time,beenakker1997random,imry1997introduction,dresselhaus1998physical,nitzan2003electron,Galperin2006Molecular,galperin2007molecular,di2008electrical,galperin2009cooling,cui2018peltier}.  
%
%
However, the cause of the limited cooling magnitude is largely unclear.
Moreover, although the influence of electron-electron interactions 
and the presence of quantum resonant states on local heating of nanojuntions
was widely studied \cite{huang2007local,ioffe2008detection,ward2011vibrational,zeng2021effect},  
how these features affect the Peltier cooling has remained a topic barely touched upon.  


To shed light on the above effects, and  
to provide new clues to enhancing Peltier cooling, 
in this Letter we investigate  
the thermoelectric response of prototypical quantum impurity systems 
to applied voltages 
by means of theoretical analysis and numerical simulations. 
Particularly, to accentuate the possible role of quantum resonance and electron correlations, 
we shall focus on 
systems which explicitly involve electron-electron interactions,
while the influence of phonons is omitted. 
We then predict an unconventional Peltier effect, we call {\it Kondo cooling}, which is the direct result of electron correlations. This phenomenon can be readily verified experimentally and offers a new 
path toward nano-refrigeration.

\begin{figure}[t]
\centering
\includegraphics[width=\columnwidth]{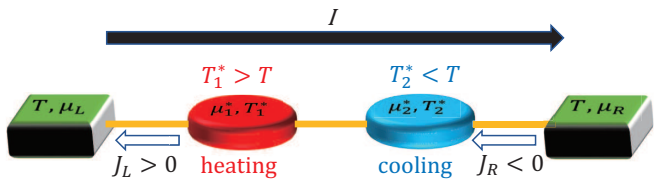}
\caption{ Schematic illustration of 
a nanojunction consisting of two localized impurities of identical energetic structure,  
which are serially connected to two leads ($L$ and $R$) at a same background temperature $T$.
With a voltage $V$ applied across the junction, the difference in electrochemical potential of the two leads 
($\mu_R - \mu_L = eV$) induces an electric current $I$ flowing through the junction. 
Because of the Peltier effect, impurity-2 has a lower local temperature than the leads
($T_2^\ast < T$), whereas impurity-1 is heated up ($T_1^\ast > T$). 
This corresponds to a net heat flow from impurity-1 to lead-$L$ ($J_L > 0$)
and from lead-$R$ to impurity-2 ($J_R < 0$), respectively. }
\label{fig1}
\end{figure}

{\it Peltier cooling with Kondo resonant states.}--
Let us first provide a 
simple analytical picture of the Kondo cooling phenomenon. Without loss of generality, the nanojunctions are represented by 
a number of localized impurities embedded between two leads. 
We first clarify that the electronic Peltier cooling cannot exist in a single-impurity junction.
This is because when an electron is driven by a bias to traverse the junction,
it gains energy from the work done by the electric field.
Such excess energy must dissipate into the surrounding (leads) in the form of heat
to conserve energy and thus sustain 
the stationary state of the junction. 
Consequently, a single impurity in the junction has no choice but to
behave as a hot spot to release the heat. 
In contrast, if there is more than one impurity in the junction, it is possible
to have one impurity cooled down by the Peltier effect, at the expense
of heating up the other impurity. 
For the sake of discussion, we then consider a junction consisting 
of two impurities connected serially to two metallic leads; see \Fig{fig1}. 

By driving a steady electric current $I$ through the junction, 
the heat current produced by the Peltier effect is 
$J^{\rm{P}}=(\Pi-\Pi_{\rm{lead}})I \approx \Pi I$ \cite{vera2016direct}, 
where $\Pi$ and $\Pi_{\rm lead}$ are the Peltier coefficients of the two-impurity complex and the leads, 
respectively, and usually $|\Pi_{\rm lead}| \ll |\Pi|$ \cite{bakker2010interplay}. 
%
Meanwhile, the impurities are also subject to Joule heating 
with the power $J^{\rm J} = IV$, which is caused by electron scattering 
within the impurities or at the impurity-lead interfaces. 
Thus, in the context of \Fig{fig1}, the net heat currents flowing into
the two leads are
\be \label{eqn:j-alpha}
 J_{L} = J_L^{\rm J}-J^{\rm P}   \quad {\rm and} \quad  J_{R} = J_R^{\rm J}+J^{\rm P},
\ee
where $J_L^{\rm J}=I(\bar{\mu}^\ast-\mu_L)/e$ and $J_R^{\rm J}=I(\mu_R-\bar{\mu}^\ast)/e$
are the Joule heat dissipated into the $L$ and $R$ leads, respectively,
with $\bar{\mu}^\ast=(\mu_1^\ast+\mu_2^\ast)/2$, and $\mu_1^\ast$ and $\mu_2^\ast$
being the local electrochemical potentials of the impurities $1$ and $2$. 
As depicted in \Fig{fig1}, the cooling of impurity-$2$ is indicated by 
a local temperature~\cite{zhang2019local,Supplementary}, $T_2^\ast$, 
lower than the background temperature, $T$, of the leads ($T_2^\ast < T$), 
as well as a reverse heat current flowing from lead-$R$ to impurity-$2$, i.e.,  
\be
  J_R = I \left(V/2 + TS\right) < 0.   \label{eqn:j-R-1}
\ee
Here, the voltage drop is assumed to be antisymmetric across the junction,
and we have used the second Thomson relation $\Pi = TS$ \cite{Rowe1983Modern,nolas2001thermoelectrics,rowe2006thermoelectrics}, with 
$S$ being the thermopower (or Seebeck coefficient) \cite{zuev2009thermoelectric,wei2009anomalous,checkelsky2009thermopower} of the two-impurity complex \cite{ziman2001electrons,wierzbicki2010electric}. 
%
Clearly, a large negative $S$ is an essential prerequisite for 
the cooling of impurity-$2$.

The zero-bias thermopower can be expressed by 
the following Landauer-like formula \cite{Don0211747,Ye2014Thermopower} 
(consider the wide-band limit for the leads and set $e = \hbar = k_{\rm B}=1$)
\be \label{S-Landauer}
S = -\frac{1}{T}\frac{\int d\omega\, (\omega-\mu)f'_\beta(\omega)A(\omega)}
{\int d\omega \, f'_\beta(\omega)A(\omega)}.
\ee
Here, $A(\w)$ is the normalized spectral density of the impurities,  
$f_\beta(\w)$ is the Fermi function, and $f'_\beta(\w)\equiv \frac{\partial f_\beta(\w)}{\partial \w}$
defines a thermal activation window,  
which is centered at the equilibrium chemical potential $\mu$
and the width is determined solely by $T$ \cite{Ye2014Thermopower}.  
The value of $S$ under a zero or low bias is thus dictated 
by the lineshape of $A(\w)$ within 
such a thermal energy window. 
%

%
In general, a quantum resonant state 
is manifested by a single spectral peak in $A(\w)$. 
From \Eq{S-Landauer}, it is clear that a sharp resonance peak 
located within the left (right) half of the thermal activation window 
will contribute a significant positive (negative) value to $S$. 
Besides, resonant states also provide ballistic channels for the 
electron transport, leading to an increased $I$. 
Therefore, quantum resonant states located slightly off the 
lead chemical potential tend to boost a reverse heat current,
and hence will greatly enhance the Peltier cooling. 

Among the various types of quantum resonant states, 
those arising due to strong electron correlations are particularly intriguing,
since a low background temperature is typically required 
to suppress the thermal fluctuations 
which impair the electron correlations.
Therefore, by exploiting the concomitant effect of Peltier cooling 
and strong electron correlations, it is possible to create a local spot 
that is cooler than the already cool surrounding.

As is well known, the screening of a localized spin moment 
by the conduction electrons in a metallic lead 
gives rise to spin-Kondo (S-Kondo) states \cite{Goldhaber1998Kondo,Cronenwett1998Tunable,Eto2000Enhancement,Pustilnik2001Kondo,Fuhrer2004Kondo,Heersche2006Kondo,Baruselli2012Kondo} at the impurity-lead interface. 
Their characteristic spectral feature 
is a sharp resonance peak centered at $\mu$.
Because of the symmetric lineshape of the spectral peak, 
while S-Kondo states enhance the junction conductance \cite{Svilans2018Thermoelectric}, 
they have little influence on the thermopower $S$ \cite{Ye2014Thermopower}. 
%

On the other hand, if the electrons are subject to 
a strong inter-impurity repulsion, 
the {\it orbital} moment of the two-impurity complex will behave like a pseudospin,
and its interaction with the leads will result in
the formation of {\it orbital-Kondo} (O-Kondo) states \cite{Jarillo2005Orbital,Choi2005SU4Kondo,Potok2007Observation,kuemmeth2008coupling,Fang2008Kondo,Karolak2011Orbital,amasha2013pseudospin}. 
%
The O-Kondo spectral signature is a pair of satellite peaks 
located on either side of $\mu$ (see also below). 
Based on \Eq{S-Landauer}, the value of $S$ can then be significantly varied 
by adjusting the relative heights of these two satellite peaks,  
which can be realized 
by tuning the energetic structures of the impurities \cite{Ye2014Thermopower}. 

The above analysis predicts unambiguously
that further cooling of an already cool impurity can be realized 
by invoking O-Kondo states in the junction. 
In the following, we shall validate such a prediction 
by performing accurate numerical simulations
on a prototypical quantum impurity model.

{\it Model and Methodology}.-- 
The total Hamiltonian 
of the junction assumes the form of an Anderson impurity model (AIM) \cite{anderson1961localized},
$\hat H = \hat H_{\rm imp}+ \hat H_{\rm lead} + \hat H_{\rm int}$.  
Specifically, the two impurities are described by 
$\hat H_{\rm imp} = \sum_{\nu=1,2} (\epsilon_0 \hat{n}_\nu + 
U \hat{n}_{\nu\uparrow} \hat{n}_{\nu\downarrow}) + U_{12} \hat{n}_{1} \hat{n}_{2}+ 
[t\,(\hat{a}_{1 \uparrow}^{\dag} \hat{a}_{2 \uparrow} + 
\hat{a}_{1 \downarrow}^{\dag} \hat{a}_{2 \downarrow}) + {\rm H.c.}]$,
where $\hat{n}_\nu = \sum_s \hat{n}_{\nu s} = \sum_s \hat{a}^\dag_{\nu s} \hat{a}_{\nu s}$, 
with $\hat{a}^\dag_{\nu s}$ ($\hat{a}_{\nu s}$) creating (annihilating) 
a spin-$s$ electron on impurity-$\nu$;
$U$ and $U_{12}$ are the intra- and inter-impurity 
electron-electron interaction energies, respectively;
and $t$ is the hopping integral between the two impurities. 
$\hat H_{\rm lead}=\sum_{\alpha ks} \epsilon_{\alpha k}\, \hat{d}^\dag_{\alpha k s} \hat{d}_{\alpha k s}$
represents the metallic leads, 
with $\hat{d}^\dag_{\alpha k s}$ ($\hat{d}_{\alpha k s}$) being the
creation (annihilation) operator for 
orbital-$k$ of lead-$\alpha$ ($\alpha=L,R$); and
$\hat H_{\rm int}=\sum_{\alpha k s \nu} t_{\alpha k \nu} \, \hat{a}^\dag_{\nu s} \hat{d}_{\alpha k s} + {\rm H.c.}$ 
describes the electron transfer couplings between the impurities and the leads,
with $t_{\alpha k \nu}$ being the coupling strengths. 
%

The thermoelectric properties 
of the junction are determined 
by a hierarchical equations of motion (HEOM) method    \cite{Tanimura1989Time,jin2008exact,li2012hierarchical,Ye2016HEOM,Han2018On,Cui2019Highly,Zhang2020Hierarchical},
which has been widely adopted to investigate strongly correlated quantum impurity systems out of equilibrium    \cite{zheng2013kondo,Ye2016HEOM,Wang2018Precise,li2020molecular,li2021reaction}. 
%
In the framework of HEOM,
the influence of the metallic leads are fully 
captured by hybridization functions 
which assume a Lorentzian form of 
$\Delta_{\alpha \nu}(\omega) \equiv \pi \sum_{k} |t_{\alpha k \nu}|^2 \delta(\w -\epsilon_{\alpha k})
=\frac{\Gamma_{\alpha \nu} W^2}{(\omega-\mu_\alpha)^2 + W^2}$,
with $\Gamma_{\alpha \nu}$ being the hybridization strength and $W$ the band width of the leads. 
%
%
The results of HEOM are numerically exact provided that
they fully converge with respect to the truncation of the hierarchy \cite{Tanimura2020Numerically}. 

\begin{figure}[t]
\centering
\includegraphics[width=\columnwidth]{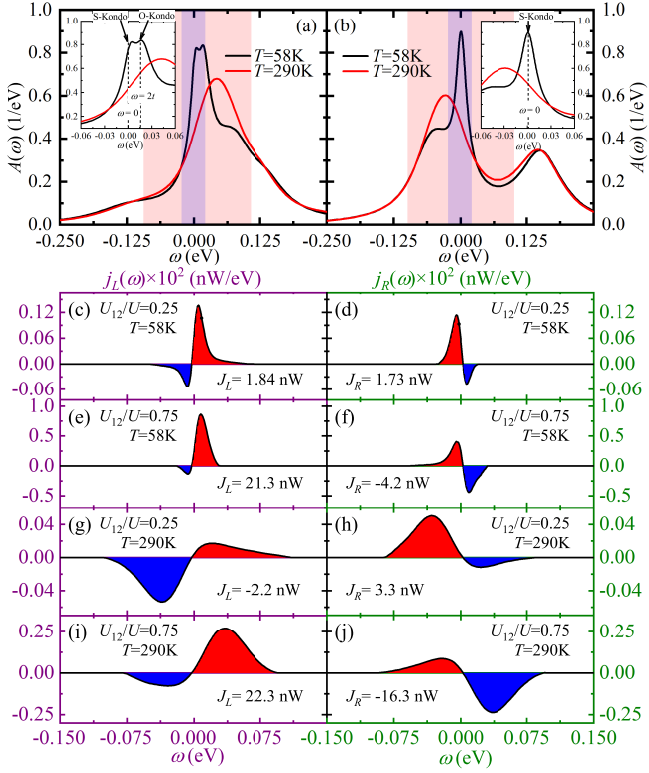}   
\caption{Normalized spectral function $A(\omega)$ of the two impurities  
with (a) $U_{12}/U=0.75$ and (b) $0.25$
under zero bias and different background temperatures.   
The areas shaded in purple and pink mark the thermal activation windows 
under $T = 58$\,K and $290$\,K, respectively. 
The insets magnify the S-Kondo and O-Kondo resonance peaks around $\mu = 0$. 
Panels (c)-(j) depict the heat current spectra $j_\alpha(\w)$
under an antisymmetric bias $\mu_R=-\mu_L=V/2=0.0025$ eV.
The energetic parameters of the AIM
are (in units of eV): $U = -2\ep_0 = 0.2$, 
$t=0.0075$, $\Gamma_{L1}=\Gamma_{R2}= 0.02$, $\Gamma_{L2}=\Gamma_{R1}=0$, and $W =1.25$. 
}\label{fig2}
\end{figure}

The realization of 
local cooling at impurity-$\nu$ is characterized by two criteria: 
(i) a reverse (negative) heat current flowing 
into impurity-$\nu$ 
from lead-$\alpha$, $J_\alpha < 0$, whose magnitude represents the power of Peltier cooling;
and (ii) 
a lower local temperature $T^\ast_\nu$ than the background temperature, 
as represented by a negative ratio $\gamma_\nu \equiv (T^\ast_\nu - T)/T < 0$. 
%
%
For the former, 
the energy spectrum of heat current $j_\alpha(\w)$ 
is calculated via the Landauer formula \cite{meir1992landauer,Supplementary}    
and $J_\alpha = \int j_\alpha(\w)\, d\w$. 
%
For the latter, 
$T_\nu^\ast$ is determined by an operational protocol based on 
a minimal perturbation condition (MPC) \cite{zhang2019local,ye2015local,ye2016thermodynamic,zeng2021effect,Supplementary},
%
%
and the values agree closely to those given by other definitions \cite{bergfield2013probing,bergfield2015tunable,shastry2016temperature,shastry2020scanning}
in the low bias region (see also Supplementary Material).


{\it Numerical demonstration of Kondo cooling.}--    
Figure~\ref{fig2}(a) depicts the spectral features of Kondo resonant states 
in the presence of a strong inter-impurity interaction ($U_{12}/U=0.75$).
Under a cryogenic temperature of $T=58\,$K,    
an S-Kondo peak shows up at $\w=\mu \equiv 0$.
Meanwhile, an O-Kondo peak emerges at $\w= 2t$, whereas its opposite counterpart at $\w = - 2t$ is invisible. 
Based on 
\Eqs{eqn:j-R-1} and \eqref{S-Landauer}, 
the asymmetric lineshape of $A(\w)$ 
around $\mu$ will give rise to a large negative $S$, 
and hence a reverse heat current from lead-$R$ to impurity-$2$, i.e., $J_R < 0$.
It is thus predicted that impurity-$2$ is cooled down below the low background $T$.  
In contrast, with a much weaker inter-impurity interaction ($U_{12}/U=0.25$), 
the O-Kondo signature vanishes almost completely; see \Fig{fig2}(b).
The resulting symmetric spectral function around $\mu$ 
will lead to a rather small thermopower $S$.
Consequently, the Peltier cooling is overwhelmed by Joule heating 
at the both impurities, i.e., $J_L > 0$ and $J_R > 0$. 

\begin{figure}[t]
\centering
\includegraphics[width=\columnwidth]{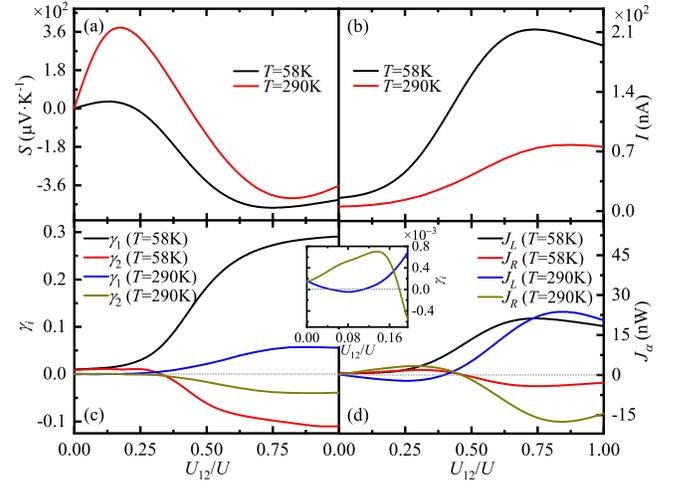}
\caption{Evolution of thermoelectric and transport properties
with respect to the variation in $U_{12}/U$: (a) thermopower $S$, (b) electric current $I$, 
(c) local temperature fluctuation $\gamma_\nu$, and (d) heat currents $J_\alpha$. 
Both cryogenic and room temperatures ($T = 58$\,K and 290\,K) are explored.
The inset magnifies the evolution of $\gamma_\nu$ in the small $U_{12}$ region. 
The energetic parameters of the AIM are the same as those adopted for \Fig{fig2}. 
}\label{fig3}
\end{figure}

At room temperature ($T = 290$\,K), 
the Kondo states are completely destroyed.
The thermopower $S$ is instead dominated by the 
non-Kondo resonant state 
whose spectral signature 
is a resonance peak located at $\w \approx \ep_0 + U_{12}$.
Since the peak center shifts from negative to positive energy 
as $U_{12}$ strengthens,  
$S$ will vary from a positive maximum to a negative maximum,    
leading to the alternation of local cooling site
from impurity-$1$ (negative $J_L$)  
to impurity-$2$ (negative $J_R$).  
%

The above prediction on the values of $S$ and the signs of $J_\alpha$ is 
validated 
by the calculated nonequilibrium transport properties.   
Figure~\ref{fig2}(c)-(j) display 
$j_\alpha(\w)$ under an antisymmetric bias.  
Under $T = 58$\,K,
$j_\alpha(\w)$ distribute tightly around zero energy,  
where Kondo resonant states contribute predominantly to 
the electron and heat transport.  
Particularly, in the case of a strong $U_{12}$,   
negative $j_R(\w)$ is promoted in the energy range 
spanned by the O-Kondo spectral peak; see \Fig{fig2}(f). 
In contrast, under $T = 290$\,K,  
$j_\alpha(\w)$ distributes over a broader energy range,
with large negative values emerging at around $\w \approx \ep_0 + U_{12}$. 
%

Figure~\ref{fig3} gives an overview of how the Peltier cooling 
is enabled by varying $U_{12}$.
Under a cryogenic temperature, a strengthened $U_{12}$
leads to an enhanced O-Kondo resonance, which in turn 
elevates the thermopower and conductance of the junction.  
The resultant Kondo cooling of impurity-$2$ is confirmed by the
fact that both the two criteria, $\gamma_2 < 0$
and $J_R < 0$, are met with a sufficiently large $U_{12}$; see \Fig{fig3}(c) and (d). 
Since the MPC-based protocol 
for the determination of $T_2^\ast$
does not require any information on heat currents, 
$\gamma_2$ and $J_R$ are calculated independently. 
It is thus remarkable that the variation of $\gamma_2$ with respect to $U_{12}$
exhibits a very similar trend to that of $J_R$,
albeit with a slightly different turning point 
from heating to cooling.


Local cooling also happens when the Kondo states are all quenched at room temperature. 
Figure~\ref{fig3}(c) and (d) reveal the same trend with 
impurity-1 subject to marginal cooling with a weak $U_{12}$, 
while the cooling site shifts to impurity-2 with a sufficiently strong $U_{12}$. 
This verifies the above prediction on the alternation 
of the local cooling site.

Unlike Kondo resonant states whose energies are 
intrinsically pinned to the leads' chemical potential, 
the positions of non-Kondo resonance peaks depend sensitively on 
the molecular orbital energies of the embedded impurities. 
%
Therefore, to have prominent non-Kondo cooling in practical applications, 
it is essential to shift the resonance energies,
such as the energy of the highest occupied molecular orbital (HOMO)
or lowest unoccupied molecular orbital (LUMO) of a molecular device,  
into the thermal activation window. 
%
In contrast, the key to realizing conspicuous Kondo cooling
is to produce strong O-Kondo resonant states 
which distribute asymmetrically with respect to 
the leads' chemical potential.


\begin{figure}[t]
\centering
\includegraphics[width=\columnwidth]{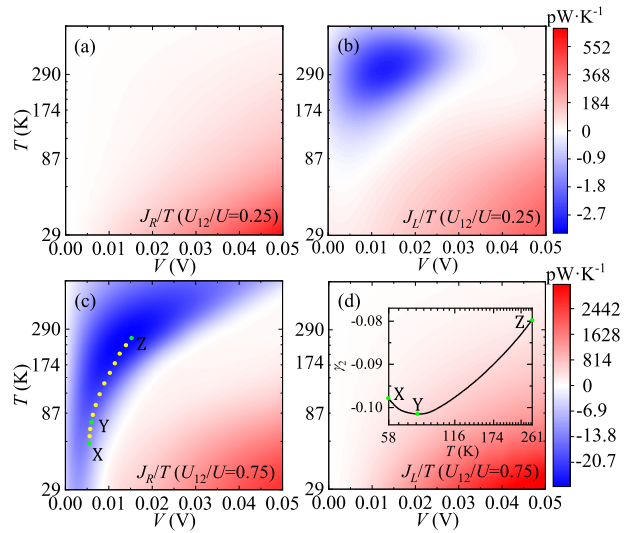}
\caption{
(a) $J_R/T$ and (b) $J_L/T$ as a function of $T$ and $V$
with $U_{12}/U = 0.25$; and
(c) $J_R/T$ and (d) $J_L/T$ as a function of $T$ and $V$
with $U_{12}/U = 0.75$.  
The voltage is applied antisymmetrically across the junction, 
i.e., $\mu_R=-\mu_L=V/2$. 
Blue (red) denotes a negative (positive) heat current
flowing out of (into) lead-$\alpha$, 
indicating local cooling (heating) of the coupled impurity. %
The inset of (d) depicts the variation of $\gamma_2$ along a path marked in (c)
which connects the Kondo and non-Kondo cooling regions.
The energetic parameters of the AIM are same as those adopted for \Fig{fig2}. 
}\label{fig4}
\end{figure}

To explore the full range of external conditions under which the Peltier cooling occurs,
we plot in \Fig{fig4} the heat currents scaled by the thermal energy, 
$J_\alpha/T$, as a function of $T$ and $V$. 
Clearly, the cooling regions (shaded in blue) exhibit
distinct patterns in the cases of weak and strong $U_{12}$. 
In the former case, local cooling takes effect 
at impurity-$1$ (corresponding to a positive $S$)
within a rather narrow range of $T$ and $V$; see \Fig{fig4}(b).
There, the cooling arises entirely from non-Kondo resonance. 
The absence of Kondo cooling under a low $T$ is due to the inactive O-Kondo effect.  
In contrast, with a strong $U_{12}$, 
local cooling exists at impurity-$2$ within a wide range of $T$ and $V$; see \Fig{fig4}(c).
Nevertheless, Kondo cooling comes into play only in the lower-left corner (low $T$ and low $V$), 
since either the intensified thermal fluctuations
or the increased mismatch between $\mu_L$ and $\mu_R$
will quench the Kondo states \cite{Meir1993Low,Franceschi2002Out,Altland2009Nonequilibrium,Cohen2014Green,Antipov2016Voltage}.  

The crossover from Kondo cooling to non-Kondo cooling is
indicated by a path marked in \Fig{fig4}(c) (the dotted line),
along which the variation of $\gamma_2$ is depicted in the inset of \Fig{fig4}(d).
It is remarkable that the Kondo resonance actually gives rise to an even
greater relative drop of local temperature than the non-Kondo resonance. 
Figure~\ref{fig4} thus highlights the important role of O-Kondo resonant states,  
as they are uniquely efficacious in further cooling
of an already cool spot 
to produce an appreciable cooling power.  

{\it Concluding remarks}.-- 
In summary, 
we have predicted an unconventional Peltier cooling phenomenon, {\it Kondo cooling}, in 
quantum impurity systems. It arises due to the crucial roles of electron-electron interactions
and quantum resonances. 
We have validated such a prediction 
numerically on a model junction. 
The existence of such prominent Kondo cooling is clearly 
indicated by the reverse heat current and a lowered local temperature, 
which originate from the {\it orbital-Kondo} resonant states distributed
unevenly around the chemical potential.
This unconventional Kondo cooling phenomenon offers both a new route toward nano-refrigeration and  
further reveals the importance of strong electron correlations 
in nonequilibrium quantum systems.

%
%


\begin{acknowledgments}
R.X. and X.Z. acknowledge the support from 
the National Natural Science Foundation of China (Grant Nos. 21973086 and 21633006), 
the Ministry of Education of China (111 Project Grant No. B18051),
and the Fundamental Research Funds for the Central Universities (Grant No. WK2060000018). 
The computational resources are provided by the Supercomputing Center 
of University of Science and Technology of China.

\end{acknowledgments}

%

\end{document}